\documentclass[aps,prl,preprint,groupedaddress]{revtex4}
\usepackage{graphics}

\begin{document}


\title{Observation of spin-bottleneck  due to spin-charge separation in a superconductor}


\author{B. Leridon}
\email[]{Brigitte.Leridon@espci.fr}
\author{J. Lesueur and M. Aprili} 
\affiliation{Laboratoire de Physique Quantique - ESPCI/CNRS - 10, rue Vauquelin -  75005  Paris - France}


\date{\today}

\begin{abstract}
 An experimental device was designed to measure the effect of the injection of spin-polarized carriers on the superconductive gap and density-of-states (DOS). Quasiparticles were injected  from a ferromagnet ($Ni_{0.8}Fe_{0.2}$) through a tunnel junction into a conventional superconductor (Nb), while charge neutrality was maintained by a supercurrent.  The DOS of the superconductor was measured  through a second tunnel junction with a normal paramagnetic metal.  No significant decrease of the superconductive gap was observed while a noticeable heating of the quasiparticles of the superconductor was measured. A similar experiment performed  with current injected from a paramagnet (Al or Ag) showed no heating of quasiparticles.   
 These observations are consistent with spin-charge separation of Bogoliubov quasi-particles and  spin-bottleneck due to the enhanced recombination time of pure spin-excitations.
 \end{abstract}


\maketitle


\textit{Introduction.} We have designed an experiment to probe the density of states of a superconductor when current is injected from a ferromagnet through a tunnel junction. 
When an electron is injected at the gap energy into a superconductor, its spin goes to the corresponding excitation, while its charge goes entirely to the condensate.   Therefore at the gap edge a Bogoliubov quasiparticle carries only spin and no charge. This point has been reemphasized by \cite{Kivelson:1990,Zhao:1995}. 
At high energies compared to the gap, however, a quasiparticle in a superconductor is similar to a metallic excitation and carries both spin and charge.  At such energies, an entering electron  transfers no charge to the condensate but as the quasiparticle relaxes towards the energy gap, its charge is  transferred to the superfluid without change in spin.  The properties of the superfluid enable the charge to be evacuated from near the injection region without any resistance. The spin excitations at the gap edge are therefore accumulated in the superconductor near the injection junction and spatially decoupled from the charge, provided their recombination rate be sufficiently large.  In the present experiment, these pure-spin excitations are also spin-polarized.  This spin polarization relaxes over a time  $\tau_{sr}$, equal to infinity if no magnetic impurities and no spin-orbit are present. In reality, if $\tau_{sr}$ is only larger than the recombination time $\tau_{r}$, then the recombination process is controlled by  $\tau_{sr}$ and the existence of  pure-spin excitations can be favoured. 
 The present paper realizes this experimental situation. Spin-polarized quasiparticles are injected from a ferromagnet ($Ni_{0.8}Fe_{0.2}$)  into the superconductor (Nb).   The charge is evacuated through the superconducting condensate by the supercurrent while the spin excitations  diffuse within a limited volume in which a second tunnel junction  acts  as an independent detector.
 
\begin{figure}
\begin{center} 
\includegraphics{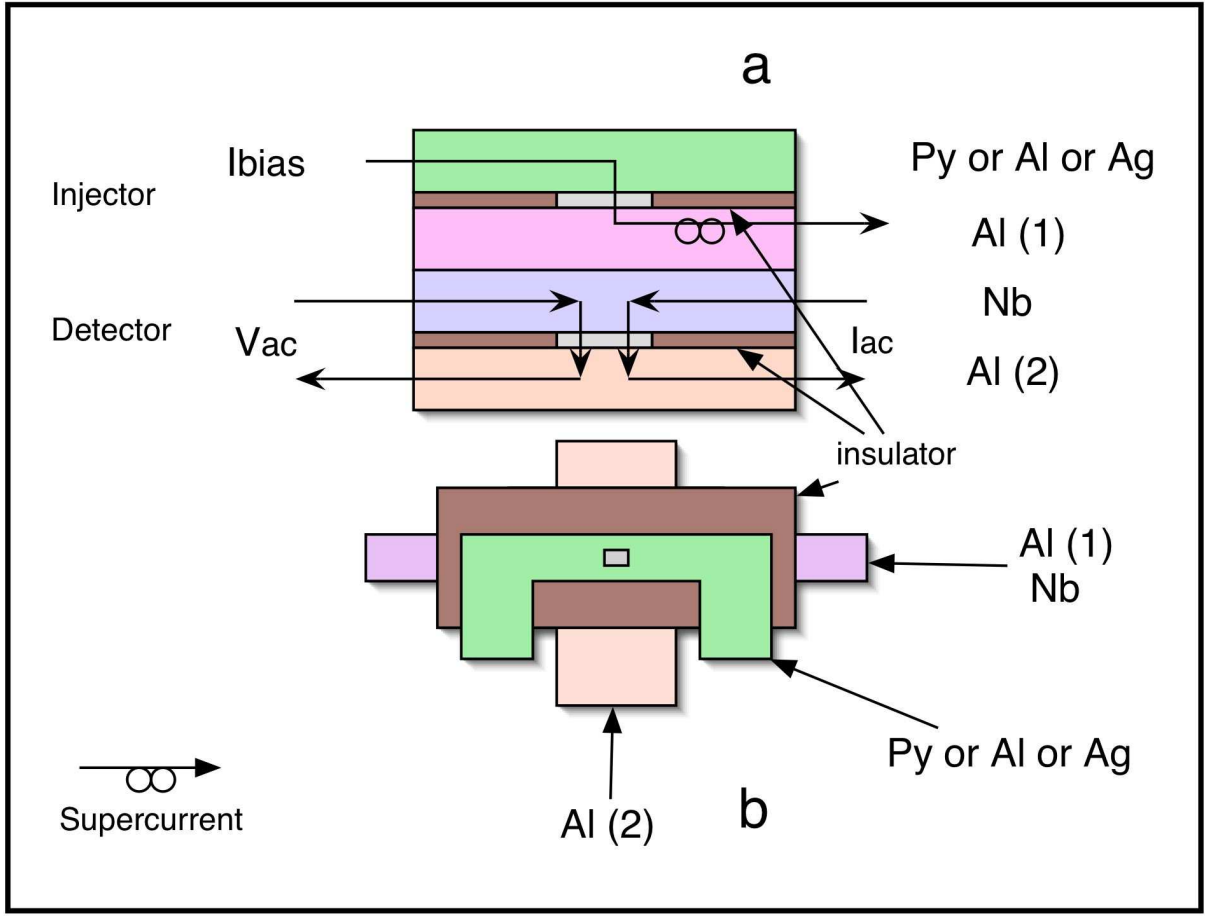}
\caption{Sketch of the device. (Color on line.)\label{fig1}}
\label{default}
\end{center}
\end{figure}

 
\textit{Experimental set-up.}  Fig.~\ref{fig1}a shows a schematic side view of the device while Fig.~\ref{fig1}b is a top view. All layers were deposited  \textit{in situ} using electron gun evaporation at  about $10^{-8}$ torr, on a Si substrate. Different patterns were defined for the metal and oxide layers by use of mechanical masks.

 The "detector" junction at the bottom of the device is a $Al/Al_2O_3/Nb$ junction of lateral dimension about  $750  ~\mu$m $\times 750  ~ \mu $m  and resistance of about 100  $\Omega$.  The superconducting layer consists of a  50 $nm$ thick Nb layer ($T_{C} = 8.2 $ K) , on top of which is deposited a 50 $nm$ thick Al layer, which is experimentally shown to be superconducting by proximity to the Nb. 
 The top junction is the "injector". Its dimensions are typically $750~\mu$m$\times750~\mu$m and its resistance at high energy is about 1 $\Omega$. The oxide barrier is also made of $Al_20_3$ realized under a shorter plasma oxidation than for the detector, and the upper layer is either a ferromagnet ($Ni_{0.8}Fe_{0.2}$), or a paramagnet (Al or Ag) for comparison.  The injector can be biased with a dc current varying from -25 mA to 25 mA. 
 
\begin{figure}
\begin{center}
\includegraphics{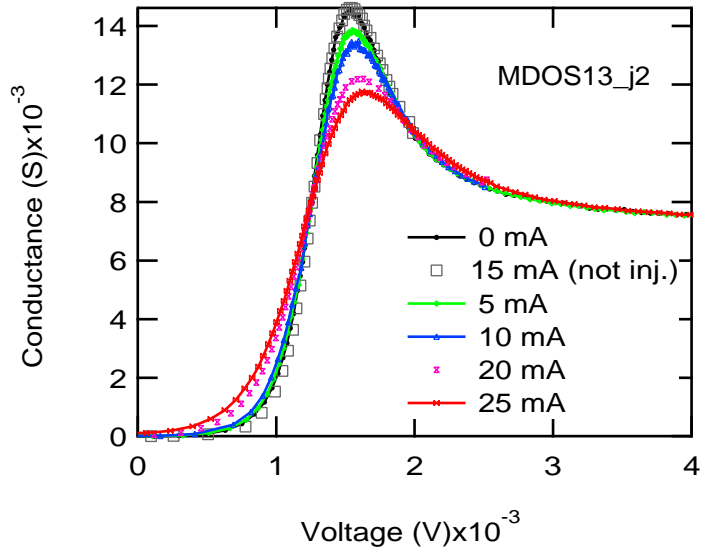}
\caption{dI/dV curves of the detector in the case of ferromagnet injection, for different values of the bias current through the injector. $T_{bath}=  1.48 $ K.  The spectrum is displayed only for  positive voltages but was symetric for negative voltages. The open grey squares correspond to a situation where the current flows through the ferromagnet layer but not through the junction, allowing to rule out spurious Joule effects through the circuitry.\label{fig2}}
\end{center}
\end{figure}

\begin{figure}
\begin{center}
\includegraphics{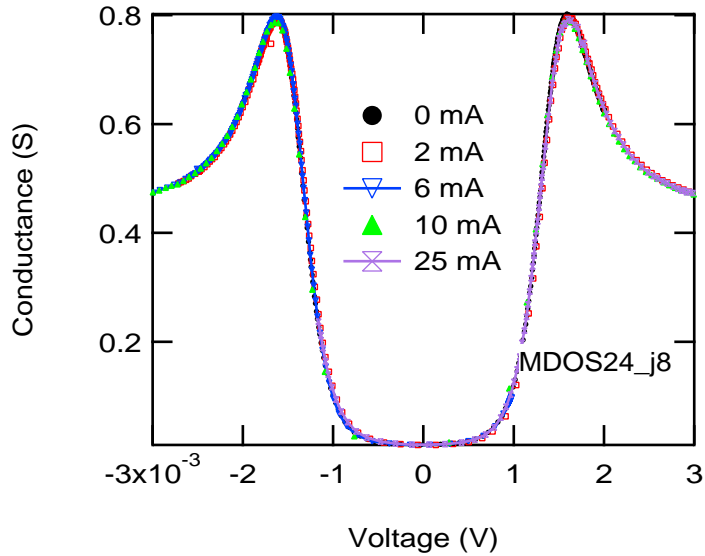}
\caption{dI/dV curves of the detector in the case of paramagnet injection, for different values of the injection current.  All curves are superimposed. $T_{bath} =  1.7$ K\label{fig3}}
\end{center}
\end{figure}

 The ac conductance of the detector was measured as a function of the dc bias voltage at about 1.5 K in a pumped helium bath dewar, for different values of the injection current. In the case when the injection is made through a \textit{ ferromagnet } injector, the spectra appear to be strongly modified (see Fig.~\ref{fig2}).  
However, when the current is applied through the \textit{paramagnet} injector into the superconductor, no effect is seen on the dI/dV curves of the detector (see Fig.~\ref{fig3}). 

\begin{figure}
\includegraphics{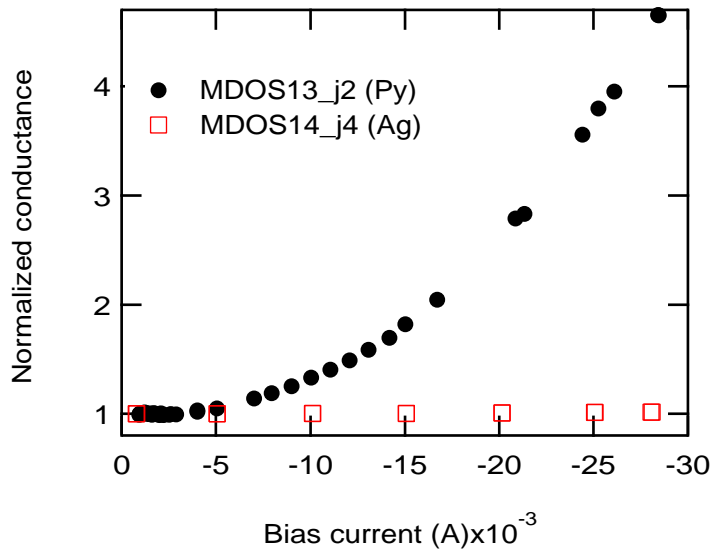}
 \caption{Variation of dI/dV at zero voltage of the detector for a ferromagnet injector (black dots) and a paramagnet injector (open red squares).  The conductance was normalized to its zero bias current value. The resistances of the injector junctions are respectively 2.6 $\Omega$ and 1.5 $\Omega$\label{fig4}}
\end{figure}

 Fig.~\ref{fig4}  illustrates a typical comparison  for a ferromagnet injection with respect to a paramagnet injection, where the conductance at zero energy is being mesured as a function of the bias current. Whereas in Fig.~\ref{fig4} no effect is visible for the paramagnet injection, actually a tiny effect  of increase of the zero voltage conductance (and correlated decrease of the gap voltage conductance) was measured with a magnitude of $1\%$ of variation at 25 mA, i.e.  400 times smaller than in the ferromagnet case.
  
These tests were made on several different devices with either ferromagnet or paramagnet injectors. These  have slightly different values of the tunnel junction resistance.  However, the existence or absence of the effect was never correlated to the values of this resistance. 
In order to definitely rule out any thermal heating effect, due to Joule effect generated through the injector resistance a test experiment was carried.  The sample with a ferromagnet injector was cooled down slightly below the critical temperature of the Al electrode of the detector (to 1.3 K).  A current was applied through the injector junction but instead of measuring the dI/dV curve of the detector we rather measured  the resistance of the Al strip.  Any local increase of the temperature of more than 50 mK due to a heating of the device over the width of the Al strip should result in non-zero resistance .  No such effect was observed for injection currents up to 20 mA.


 \textit{Data analysis. } The principal observation is the considerable difference between the spectra obtained on the junctions where the injector is a ferromagnet or a paramagnet.  The spectra are shown to be strongly dependent on the injected spin-polarized current.   Actually they are found to be very well fitted by a conventional BCS tunneling expression with a temperature higher than the He-bath temperature. One may recall here that an equilibrium BCS junction constitutes by itself an idealized thermometer for the electrons. Once the conductance is normalized to its value at high temperature and the gap to its zero-temperature value, the curve depends only on the temperature of the metal.  It is quite remarkable that such a simple result is being obtained.   The gap amplitude shows only a slight decrease. This indicates that injected itinerant electrons do behave quite differently from localized magnetic moments. 
 
  The expression giving the conductance through a tunnel junction in the framework of BCS theory is the following~\cite{Tinkham:1972}, but please see
  \footnote{The general expression for the current derived in ~\cite{Tinkham:1972} holds as well for polarized quasiparticles:
  $I(V)=\frac{4\pi e}{\hbar}|T|^{2}D_{N}(E_{F})\times\int_{\Delta}^{\infty}{\{[f_{k<}-f_{k>}]+D_{S}(E)\times[f(E-eV)-f(E+eV)]\}dE}$ 
, where $ f_{k>}$ and $f_{k<}$ refer to the out-of-equilibrium occupation distributions for the quasi-electron and quasi-hole branches respectively in the superconductor.  The first term contribution does not carry any energy dependence, is expected to be small and is not able to account for the experimental observation.}
 
  \begin{equation}
G(V)=-\frac{4\pi e^{2}}{\hbar}|T|^{2}D_{N}(E_{F})\int_{0}^{\infty}{D_{S}(E)\frac{\partial{f}}{\partial{E}}(E+eV)dE}
\end{equation}

, where $D_{S}(E)$ is the  density of states of the superconductor and $f(E)$ refers to the Fermi-Dirac distribution on the metal side.
  
\begin{figure}
\includegraphics{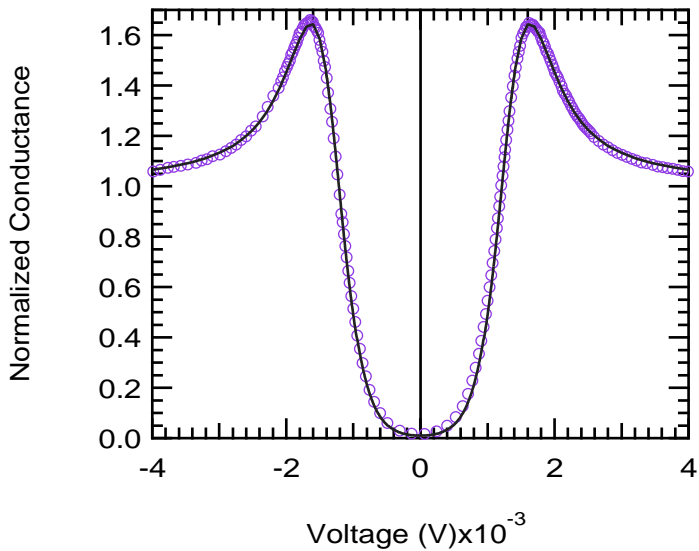}
\caption{Fit using BCS theory of the spectrum measured under a spin-polarized current of 25 mA. circles: experimental data; line, BCS fit using $T^{\ast}=2.38$ K and $ \Delta=1.37$ meV.\label{fig5}}
\end{figure}

  The parameters for the fits are given on Table 1. An example of fit corresponding to a spin-polarized current of 25 mA is shown on Fig.~\ref{fig5}.

 \textit{Interpretation. } The net effect of injecting a spin-polarized current is observed to be an enhanced  electronic temperature of the metal. The fact that under otherwise identical conditions, the current injected from a paramagnet produced no comparable change in the I -V curves rules out simple Joule heating as an explanation.
 
  It has been observed that any massive injection of quasiparticles in a superconductor leads to an out-of-equilibrium state where excitations may remain at a high temperature, thermalized with high-energy phonons (phonons with energies greater than $2\triangle{}$ ), while the condensate remains in equilibrium with low energy phonons ~\cite{Parker:1975}.  This imbalance is controlled by the injection rate of quasiparticles (and to a certain extent by their energies when entering the superconductor) and by the recombination time of the quasiparticles $\tau_{r}$.  The phonon escape time may also play a role and contribute to a greater effective $ \tau_{r}$. Similar out-of-equibrium effects \textit{without}  spin-polarization were very carefully studied experimentally by different groups ~\cite{Ginzberg:1962,Hu:1974,Sai-Halasz:1974,Jaworski:1979,Johnson:1991} and theoretically by ~\cite{Owen:1972,Chang:1974,Parker:1975}. 
 
 In the case of spin-polarized injection, the actual $ \tau_{r}$  is substantially modified for a majority of the excitations due to their spin polarization. Two quasiparticles with the same spin-polarization can not recombine to the condensate.  Therefore, if the excitations of the superconductor have a polarization rate of $P$, then the fraction $2(1-P)$ recombine over a time $\tau_{r}$, and the fraction $2P-1$ must experience spin-flip before recombining, i.e. will recombine over $\tau_{r}+\tau_{sr}$. If the spin relaxation time $\tau_{sr}$ is much greater than $ \tau_{r}$ and the inelastic scattering time $\tau_s$, they come to equilibrium at a temperature $T*$ and a chemical potential $\mu*$ different from that of the condensate.  The "bottle-neck" is then the spin relaxation. 
   
    The most likely source of spin polarization relaxation in Nb, is spin-orbit scattering from heavier impurities (like Ta). This has been measured many years ago \cite{Vier:1983} in a limited temperature range near $T_c$ and a theory was provided by Yafet \cite{Yafet:1983}. The relaxation time diverges at low T because of the BCS coherence factor, which takes into account that the matrix element for spin-orbit scattering renormalizes to zero at the gap edge  because the excitations  at such energy do not carry charge. Spin-charge separation therefore modifies the spin dynamics in the SC state and induces a larger $\tau_{sr}$.
  $\tau_{sr}$ will vary from sample to sample but taking the measurements \cite{Vier:1983}, and extrapolating using the theory by Yafet, we estimate $\tau_{sr}$ at 2~K in Nb to be about $10^{-8}$ s. 
  Johnson \cite{Johnson:1991} has measured $\tau_{r}$ in Nb down to $T/T_c =0.78$ and finds it at the lowest temperature to be  about $10^{-11} $ s. Using this value and extrapolating, following Kaplan et al. \cite{Kaplan:1976} to a temperature of 2~K, $\tau_{r}$ is estimated to be about $10^{-9}$ s.  It may actually be even smaller because  what is calculated in Kaplan's work is the $\tau_{r}$ for two quasiparticles, the rest of the particles being at equilibrium. Seminozhenko \cite{Seminozhenko:1974} has shown that in the case of an out-of-equilibrium distribution $\tau_r$ is further reduced.
 
 $\tau_{s}$ has been calculated also for a quasiparticle at a given energy, the lattice being at equilibrium \cite{Kaplan:1976}.  For a particle injected at energy $2 \Delta$, $\tau_{s}$ is about $ 6\times10^{-11} $~s. Of course it varies with the energy of the injected quasiparticle and has not been calculated for the present case of a high density of quasiparticles at a high effective temperature together with hot thermalized phonons. In our case, the thermalization may also be determined by particle-particle scattering\cite{Pierre:1999}. The fact that our tunneling curves fit a "thermal distribution" is an evidence that the thermalization time is substantially shorter than the spin-flip blocked recombination time. 
 Parker~\cite{Parker:1975} has previously observed the effects of an increased quasiparticle density in an experiment in which the bottleneck is the phonon escape time rather than the spin-flip time as in our case.  We may estimate, using his procedure that the density of excess quasiparticles for an injection current of 20 mA is about  80 times the expected number at the bath temperature. 
The out-of-equibrium excitations adopt a thermal distribution characterized by a temperature $T^{\ast}$and a chemical potential $\mu^{\ast}$. We may estimate the change $\mu^{\ast}-\mu$ to be of the order of the chemical potential shift observed in charge-imbalance experiments (less than 1 nV), consistent with its lack of importance in our experiment compared to that of $T^{\ast}$.
 
 Our experiment shows through the fit to Eq. (1) shown in Fig.~\ref{fig5}, that the detector electrode Al at least in the vicinity of the tunneling region also comes to thermal equilibrium with the thermalized  spin-only quasiparticles of Nb. Our experiment does not provide the mechanism for this process. The equilibration  mechanism may be diffusion of the "hot" phonons of Nb across the barrier, inelastic scattering with phonons or  electron-electron interactions\cite{Pierre:1999}.     All these processes are   helped by the low heat capacity of the Al. \footnote{The power delivered to the quasiparticles in the injector (about $10^{-4}$ W for I=10 mA) must roughly equal the power needed to ensure a gradient of electronic temperature of about 1~K over the Al(2) strip width of 1mm. This is met for an estimated value of the thermal conductivity of 10 $W.cm.K^{-1}$ in Al.}
 

\textit{Conclusion. }We injected tunneling particles from a ferromagnet into a superconductor. The quasiparticles thermalize quickly to near the gap edge where the incoming charge goes into the condensate. This charge is drained by a supercurrent leaving a  distribution of zero charge, spin-only (at the gap edge) quasiparticles. Because they carry no charge, these have a reduced spin-orbit scattering and therefore a long spin-relaxation time, which increases their recombination time so that they thermalize to a temperature different from the condensate.   These consequences of the spin-charge separation are directly measured by probing the effective temperature of outgoing particles using a tunnel detector. No such rise in temperature is obtained when the same current is injected by a paramagnet rather than a ferromagnet.   
 



 \begin{table}
\caption{Results of the fits to a BCS tunneling conductance curve}
 \begin{ruledtabular}
\begin{tabular}{c | c | c | c }
 Bias current (mA) & $T_{bath} $(K) & $T^{\ast} (K)$ & $\Delta $(meV) \\
 \hline
& $\pm5$~mK &$ \pm80$~mK & $\pm10~\mu$eV \\
 0 & 4.21 & 4.21 & 1.28  \\   
 0  & 1.48 & 1.48 &  1.38\\  
 4.98 & 1.47 & 1.61 & 1.38 \\   
  9.99& 1.47 & 1.71 & 1.38 \\  
  14.72 & 1.47 & 1.75 & 1.375\\   
 19.95 & 1.47 & 2.10 & 1.37\\  
  25 & 1.48 & 2.38 & 1.37  \\  
 
 \end{tabular}
 \end{ruledtabular}
 \end{table}

 \begin{acknowledgments}
 The authors thank S. Caprara, M. Grilli, P. Monod for stimulating discussions.  One of us (BL) gratefully acknowledges  C.M. Varma  for numerous discussions and  fruitful suggestions, especially as interpreting the data in terms of spin-charge separation.  The experiments were partly supported by the french Minist\`ere de la Recherche through an "Action concert\'ee incitative".
\end{acknowledgments}

\bibliography{Mybib}

\begin{thebibliography}{16}
\expandafter\ifx\csname natexlab\endcsname\relax\def\natexlab#1{#1}\fi
\expandafter\ifx\csname bibnamefont\endcsname\relax
  \def\bibnamefont#1{#1}\fi
\expandafter\ifx\csname bibfnamefont\endcsname\relax
  \def\bibfnamefont#1{#1}\fi
\expandafter\ifx\csname citenamefont\endcsname\relax
  \def\citenamefont#1{#1}\fi
\expandafter\ifx\csname url\endcsname\relax
  \def\url#1{\texttt{#1}}\fi
\expandafter\ifx\csname urlprefix\endcsname\relax\def\urlprefix{URL }\fi
\providecommand{\bibinfo}[2]{#2}
\providecommand{\eprint}[2][]{\url{#2}}

\bibitem[{\citenamefont{Kivelson and Rokhsar}(1990)}]{Kivelson:1990}
\bibinfo{author}{\bibfnamefont{S.}~\bibnamefont{Kivelson}} \bibnamefont{and}
  \bibinfo{author}{\bibfnamefont{D.}~\bibnamefont{Rokhsar}},
  \bibinfo{journal}{Phys. Rev. B} \textbf{\bibinfo{volume}{41}},
  \bibinfo{pages}{11693} (\bibinfo{year}{1990}).

\bibitem[{\citenamefont{Zhao and Hershfield}(1995)}]{Zhao:1995}
\bibinfo{author}{\bibfnamefont{H.~L.} \bibnamefont{Zhao}} \bibnamefont{and}
  \bibinfo{author}{\bibfnamefont{S.}~\bibnamefont{Hershfield}},
  \bibinfo{journal}{Phys. Rev. B} \textbf{\bibinfo{volume}{52}},
  \bibinfo{pages}{3632} (\bibinfo{year}{1995}).

\bibitem[{\citenamefont{Tinkham}(1972)}]{Tinkham:1972}
\bibinfo{author}{\bibfnamefont{M.}~\bibnamefont{Tinkham}},
  \bibinfo{journal}{Phys. Rev. B} \textbf{\bibinfo{volume}{6}},
  \bibinfo{pages}{1747} (\bibinfo{year}{1972}).

\bibitem[{\citenamefont{Parker}(1975)}]{Parker:1975}
\bibinfo{author}{\bibfnamefont{W.~H.} \bibnamefont{Parker}},
  \bibinfo{journal}{Phys. Rev. B} \textbf{\bibinfo{volume}{12}},
  \bibinfo{pages}{3667} (\bibinfo{year}{1975}).

\bibitem[{\citenamefont{Ginzberg}(1962)}]{Ginzberg:1962}
\bibinfo{author}{\bibfnamefont{D.~M.} \bibnamefont{Ginzberg}},
  \bibinfo{journal}{Phys. Rev. Lett.} \textbf{\bibinfo{volume}{8}},
  \bibinfo{pages}{204} (\bibinfo{year}{1962}).

\bibitem[{\citenamefont{Hu et~al.}(1974)\citenamefont{Hu, Dynes, and
  Narayanamurti}}]{Hu:1974}
\bibinfo{author}{\bibfnamefont{P.}~\bibnamefont{Hu}},
  \bibinfo{author}{\bibfnamefont{R.~C.} \bibnamefont{Dynes}}, \bibnamefont{and}
  \bibinfo{author}{\bibfnamefont{V.}~\bibnamefont{Narayanamurti}},
  \bibinfo{journal}{Phys. Rev. B.} \textbf{\bibinfo{volume}{10}},
  \bibinfo{pages}{2786} (\bibinfo{year}{1974}).

\bibitem[{\citenamefont{Sai-Halasz et~al.}(1974)\citenamefont{Sai-Halasz, Chi,
  Denenstein, and Langenberg}}]{Sai-Halasz:1974}
\bibinfo{author}{\bibfnamefont{G.}~\bibnamefont{Sai-Halasz}},
  \bibinfo{author}{\bibfnamefont{C.~C.} \bibnamefont{Chi}},
  \bibinfo{author}{\bibfnamefont{A.}~\bibnamefont{Denenstein}},
  \bibnamefont{and} \bibinfo{author}{\bibfnamefont{D.~N.}
  \bibnamefont{Langenberg}}, \bibinfo{journal}{Phys. Rev. Lett.}
  \textbf{\bibinfo{volume}{33}}, \bibinfo{pages}{215} (\bibinfo{year}{1974}).

\bibitem[{\citenamefont{Jaworski and Parker}(1979)}]{Jaworski:1979}
\bibinfo{author}{\bibfnamefont{F.}~\bibnamefont{Jaworski}} \bibnamefont{and}
  \bibinfo{author}{\bibfnamefont{W.~H.} \bibnamefont{Parker}},
  \bibinfo{journal}{Phys. Rev. B.} \textbf{\bibinfo{volume}{20}},
  \bibinfo{pages}{945} (\bibinfo{year}{1979}).

\bibitem[{\citenamefont{Johnson}(1991)}]{Johnson:1991}
\bibinfo{author}{\bibfnamefont{M.}~\bibnamefont{Johnson}},
  \bibinfo{journal}{Phys. Rev. Lett.} \textbf{\bibinfo{volume}{67}},
  \bibinfo{pages}{374} (\bibinfo{year}{1991}).

\bibitem[{\citenamefont{Owen and Scalapino}(1972)}]{Owen:1972}
\bibinfo{author}{\bibfnamefont{C.~S.} \bibnamefont{Owen}} \bibnamefont{and}
  \bibinfo{author}{\bibfnamefont{D.~J.} \bibnamefont{Scalapino}},
  \bibinfo{journal}{Phys. Rev. Lett.} \textbf{\bibinfo{volume}{28}},
  \bibinfo{pages}{1559} (\bibinfo{year}{1972}).

\bibitem[{\citenamefont{Chang and Scalapino}(1974)}]{Chang:1974}
\bibinfo{author}{\bibfnamefont{J.-J.} \bibnamefont{Chang}} \bibnamefont{and}
  \bibinfo{author}{\bibfnamefont{D.~J.} \bibnamefont{Scalapino}},
  \bibinfo{journal}{Phys. Rev. B.} \textbf{\bibinfo{volume}{10}},
  \bibinfo{pages}{4047} (\bibinfo{year}{1974}).

\bibitem[{\citenamefont{Vier and Schultz}(1983)}]{Vier:1983}
\bibinfo{author}{\bibfnamefont{D.}~\bibnamefont{Vier}} \bibnamefont{and}
  \bibinfo{author}{\bibfnamefont{S.}~\bibnamefont{Schultz}},
  \bibinfo{journal}{Phys. Lett. A} \textbf{\bibinfo{volume}{98}},
  \bibinfo{pages}{283} (\bibinfo{year}{1983}).

\bibitem[{\citenamefont{Yafet}(1983)}]{Yafet:1983}
\bibinfo{author}{\bibfnamefont{Y.}~\bibnamefont{Yafet}},
  \bibinfo{journal}{Phys. Lett. A} \textbf{\bibinfo{volume}{98}},
  \bibinfo{pages}{287} (\bibinfo{year}{1983}).

\bibitem[{\citenamefont{Kaplan et~al.}(1976)\citenamefont{Kaplan, Chi,
  Langenberg, Chang, Jafarey, and Scalapino}}]{Kaplan:1976}
\bibinfo{author}{\bibfnamefont{S.~B.} \bibnamefont{Kaplan}},
  \bibinfo{author}{\bibfnamefont{C.~C.} \bibnamefont{Chi}},
  \bibinfo{author}{\bibfnamefont{D.~N.} \bibnamefont{Langenberg}},
  \bibinfo{author}{\bibfnamefont{J.-J.} \bibnamefont{Chang}},
  \bibinfo{author}{\bibfnamefont{S.}~\bibnamefont{Jafarey}}, \bibnamefont{and}
  \bibinfo{author}{\bibfnamefont{D.~J.} \bibnamefont{Scalapino}},
  \bibinfo{journal}{Phys. Rev. B} \textbf{\bibinfo{volume}{14}},
  \bibinfo{pages}{4854} (\bibinfo{year}{1976}).

\bibitem[{\citenamefont{Seminozhenko}(1974)}]{Seminozhenko:1974}
\bibinfo{author}{\bibfnamefont{V.}~\bibnamefont{Seminozhenko}},
  \bibinfo{journal}{Sov. Phys. Solid State} \textbf{\bibinfo{volume}{15}},
  \bibinfo{pages}{1945} (\bibinfo{year}{1974}).

\bibitem[{\citenamefont{Pierre et~al.}(2000)\citenamefont{Pierre, Pothier,
  Est\`eve, and Devoret}}]{Pierre:1999}
\bibinfo{author}{\bibfnamefont{F.}~\bibnamefont{Pierre}},
  \bibinfo{author}{\bibfnamefont{H.}~\bibnamefont{Pothier}},
  \bibinfo{author}{\bibfnamefont{D.}~\bibnamefont{Est\`eve}}, \bibnamefont{and}
  \bibinfo{author}{\bibfnamefont{M.}~\bibnamefont{Devoret}},
  \bibinfo{journal}{J. Low Temp. Phys.} \textbf{\bibinfo{volume}{118}},
  \bibinfo{pages}{437} (\bibinfo{year}{2000}).

\end{thebibliography}

\end{document}